\pdfoutput=1  
\documentclass[conference]{IEEEtran}

\usepackage[T1]{fontenc}
\usepackage[utf8]{inputenc}
\usepackage{amsmath,amssymb}
\usepackage{booktabs}
\usepackage{array}
\usepackage{cite}
\usepackage{url}
\usepackage{xcolor}
\usepackage{dblfloatfix}
\usepackage{pgfplots}
\pgfplotsset{compat=1.18}
\usepackage[hidelinks]{hyperref}
\usepackage{orcidlink}
\usepackage{comment}
\usepackage{latexml}
\usepackage{soul}
\newcommand{\code}[1]{\texttt{#1}}

\newif\ifarxiv
\arxivtrue          

\ifarxiv
  \newcommand{\arxivbanner}{%
    This paper has been accepted for publication at the\par
    Forum on Specification and Design Languages (FDL), Rome, Italy, 2026}
  \makeatletter
  \let\arxiv@orig@titlepagestyle\ps@IEEEtitlepagestyle
  \def\ps@IEEEtitlepagestyle{%
    \arxiv@orig@titlepagestyle
    \def\@oddhead{\hb@xt@\textwidth{\hss\raisebox{-13pt}[0pt][0pt]{%
      \parbox[b]{\textwidth}{\centering\normalsize\color{gray}\arxivbanner}}\hss}}%
    \let\@evenhead\@oddhead}
  \makeatother
  \hypersetup{%
    pdftitle={Quantifying the Effect of HCLs on a Fixed-Microarchitecture MXFP4 Accelerator},
    pdfauthor={Daniele Passaretti, Sajjad Tamimi, Nicola Dall'Ora},
    pdfsubject={Forum on Specification and Design Languages (FDL) 2026 -- Special Session on Hardware Construction Languages},
    pdfkeywords={hardware construction languages; HCL; domain-specific language; microscaling floating-point; MXFP4; accelerator design; hardware generators; FPGA; high-level synthesis}}
\fi

\begin{document}

\title{Quantifying the Effect of HCLs on a Fixed-Microarchitecture MXFP4 Accelerator}

\author{%
\IEEEauthorblockN{
Daniele Passaretti\,\orcidlink{0000-0001-7154-8354}\IEEEauthorrefmark{1},
Sajjad~Tamimi\,\orcidlink{0000-0001-8092-2969}\IEEEauthorrefmark{2},
Nicola Dall'Ora\,\orcidlink{0000-0003-0656-9786}\IEEEauthorrefmark{3}
}

\iflatexml
\IEEEauthorblockA{%
\IEEEauthorrefmark{1}Institute for Information Technology and Communications,
Otto-von-Guericke University Magdeburg, Germany\\
\IEEEauthorrefmark{2}Independent Researcher, Munich, Germany\\
\IEEEauthorrefmark{3}Department of Engineering Sciences,
Universit\`a degli Studi Guglielmo Marconi, Rome, Italy}
\else
\IEEEauthorblockA{%
\IEEEauthorrefmark{1}Institute for Information Technology and Communications,
Otto-von-Guericke University Magdeburg, Germany\\
\IEEEauthorrefmark{3}Department of Engineering Sciences,
Universit\`a degli Studi Guglielmo Marconi, Rome, Italy\\
\IEEEauthorrefmark{2}Independent Researcher, Munich, Germany}
\fi

\thanks{Corresponding author: D. Passaretti (daniele.passaretti@ovgu.de).}
}%

\IEEEspecialpapernotice{Special Session Paper}

\IEEEoverridecommandlockouts
\IEEEpubid{\makebox[\columnwidth]{
979-8-3195-3997-7/26/\$31.00~\copyright2026
IEEE \hfill} \hspace{\columnsep}\makebox[\columnwidth]{ }}

\maketitle

\begin{abstract}
Hardware Construction Languages (HCLs) aim to improve hardware design productivity while generating register-transfer-level (RTL) circuits without changing the designer’s microarchitecture. However, most comparisons between HCLs are either qualitative or evaluate quality of results (QoR) across different designs, making it difficult to separate language effects from design effects. This paper compares the most widely used HCLs using the same fixed design, the OCP MXFP4 block dot product, a quantization primitive at the heart of edge \emph{Physical-AI} inference, implemented as a single \mbox{12-stage}, \mbox{II$=$1} pipeline. A SystemVerilog baseline is followed by implementations in Chisel, SpinalHDL, Amaranth, Clash, Bluespec, and C++ for high-level synthesis (HLS). Every variant goes through the same flow on the same Artix-7 device set at 100 MhZ, driven by a RISC-V soft core. With the micro-architecture held constant, the comparison is clean: every variant meets timing, and the HCLs match or even undercut hand-written RTL in area. The remaining differences stem not from the algorithm but from how each back end lowers arithmetic, and from a single width choice that silently toggles DSP inference. Unlike HLS, where design decisions are limited to pragmas, the HCLs achieve comparable area and timing.  Therefore, the choice comes down to ecosystem fit and interface needs rather than QoR.

\end{abstract}


\begin{table*}[!t]
\centering
\caption{The seven MXFP4 dot-product variants and their post-route QoR.}
\label{tab:main}
\renewcommand{\arraystretch}{1.0}
\setlength{\tabcolsep}{5pt}
\begin{tabular}{@{}lllrrrrrrr@{}}
\toprule
\textbf{Variant} & \textbf{Base lang.} & \textbf{Paradigm} & \textbf{SLoC}
& \textbf{LUT} & \textbf{SRL} & \textbf{FF} & \textbf{DSP}
& \textbf{Slack (ns)} & \textbf{$F_{\max}$ (MHz)}\\
\midrule
Chisel        & Scala   & generator HCL   & 298 & 10846 & 1096 & 3367  & 0  & $+0.290$ & 103.0 \\
SpinalHDL     & Scala   & generator HCL   & 206 &  8602 & 1096 & 3357  & 0  & $+0.251$ & 102.6 \\
Amaranth      & Python  & generator HCL   & 255 &  9096 & 1096 & 3482  & 0  & $+0.187$ & 101.9 \\
Clash         & Haskell & functional HCL  & 290 &  7361 & 1096 & 2540  & 46 & $+0.307$ & 103.2 \\
Bluespec      & BSV     & rule-based HCL  & 281 &  7333 & 1098 & 5050  & 32 & $+0.290$ & 103.0 \\
\midrule
SystemVerilog & ---     & reference RTL   & 705 & 11025 & 1096 & 3368  & 0  & $+0.135$ & 101.4 \\
Vitis HLS$^{\dagger}$ & C++ & algorithmic HLS & 292 & 11662 & 2506 & 11059 & 0 & --- & --- \\
\bottomrule
\end{tabular}
\vspace{2pt}
\\[-2pt]
{\footnotesize SLoC counts the pipeline-description source lines, excluding
short elaboration drivers; the SystemVerilog data path is hand-written, not
generated. $^{\dagger}$HLS figures are for the standalone Vitis IP with a
streaming interface and are not directly comparable to the parallel pipeline.}
\end{table*}

\section{Introduction}
HCLs have been adopted successfully in large designs, from soft-core processors (e.g., Rocket~\cite{rocket}) to ASIC accelerator front ends and host-FPGA interfacing systems~\cite{tamimi2022fccm}. Yet many hardware designs, such as for critical edge~\cite{passaretti2022isolation,passaretti2023enabling}, are still written in conventional HDLs due to designer familiarity and predictable low-level control. The question is therefore no longer whether to use an HCL, but which one as consequential choice, since no normalization path exists between the languages. Existing comparisons provide limited guidance: they measure one language across many benchmarks, or many languages across heterogeneous designs, so language and design jointly shape the reported numbers and cannot be separated.


In this work, we separate these effects with a controlled, single-kernel experiment in which the same RTL microarchitecture is described, bit-for-bit, in five HCLs and a hand-written SystemVerilog reference. An HLS variant is included as an upper baseline, and all variants are measured after place-and-route on the same FPGA. In this study, we target the OCP Microscaling 4-bit Floating-Point format (MXFP4) and integrate it into a RISC-V soft core as an accelerator~\cite{mxspec}. Three findings emerge. The HCLs impose neither a frequency nor an area penalty over hand-written RTL. The remaining QoR variance comes from arithmetic lowering and an incidental width choice that moves the products from logic fabric to DSP slices, not from the algorithm. HLS changes the solution itself and pays for it in flip-flops.

\section{MXFP4 micro-architecture}
MXFP4 is widely used in neural-network inference: a single E8M0 scale shared across the block gives the block an FP32 exponent range, while the 4-bit E2M1 elements place each value within the block and span a ratio of approximately 12× (from 0.5 to 6) between the smallest and largest non-zero magnitude. This permits to run models near sensor  with the fraction of the memory and energy of \mbox{FP16}. Each block holds 32 \mbox{E2M1} elements (a sign and a \mbox{3-bit} magnitude) sharing one \mbox{E8M0} scale. Inputs and output are \mbox{FP32}: the design operates directly on the bit patterns, in four phases: a max-exponent tournament over the 32 FP32 fields produces the shared exponent and clamped \mbox{E8M0} scale; 64 quantizers round each operand to \mbox{E2M1} (round-to-nearest, ties to even); the decoded magnitudes are multiplied pairwise and summed in a \mbox{16-bit} adder tree; and the integer dot product is cast back to \mbox{FP32}. 
The reference micro-architecture is a \mbox{12-stage}, \mbox{II=$1$} SystemVerilog pipeline that closes at \mbox{100\,MHz}. The full operand pair, $2{\times}32{\times}32 = 2048$
 bits, is presented in parallel on every cycle; control travels through a \code{valid} shift register. For system-on-chip integration, the core is a loosely coupled accelerator attached to the MicroBlaze MCS V, a RISC-V soft core, through a thin GPIO wrapper, identical for every variant and excluded from the QoR.

\section{Implementations and Methodology}
For the analysis, we have implemented seven variants. The five HCLs and
the SystemVerilog reference all describe the same \mbox{12-stage} structure,
register for register. Only the HLS variant differs from it, expressing the
same arithmetic as four FIFO-coupled \code{DATAFLOW} processes (\mbox{II$=$1}). In the specific of HCL, we have choosen Chisel~\cite{chisel} and SpinalHDL~\cite{spinalhdl}, as Scala HCL generators, Amaranth~\cite{amaranth}, as a Python HCL generator, Clash~\cite{clash}, as Haskell functional HCL generator, and Bluespec~\cite{bluespec}, as rule-based HCL generator. HLS~\cite{hlssurvey} serves only
as an upper baseline and is not an HCL.
Each back end emits the same core, \code{MXFP4Top}. We then wrap it in the
identical \code{mxfp4\_mcs\_wrapper}, place it in the same block design, using Vivado 2025.1 and the \code{xc7a100tcsg324-1}  FPGA, and a single \mbox{100\,MHz} constraint. The numbers we
report are post-route utilization and timing for the core itself. The wrapper
adds the same overhead to every variant.

\section{Results and Discussion}
Table~\ref{tab:main} reports the post-route QoR, from which four observations
stand out. First, timing does not discriminate: every variant closes at
\mbox{100\,MHz} with positive slack, and $F_{\max}$ stays within a narrow
$101$--$103$\,MHz band. The design is bound by area, not frequency, so the
variants separate on resources alone.

On resources, the HCL abstraction turns out to be almost free. The
SystemVerilog reference has the highest LUT count (\mbox{11.0k}) and the
tightest slack; every HCL matches or undercuts it with a fraction of the code,
and Clash and Bluespec are the smallest of all. On a structured data path, the
higher-level description costs nothing in QoR.

However, a single type width silently flips
DSP inference. In fact, Clash and Bluespec map the 32 element products onto DSP48
slices, while the other four keep them in LUT and CARRY logic, as forced in
the hand-written RTL. The cause is the declared width of the decoded
\mbox{E2M1} value: Clash and Bluespec declare it at \mbox{16 bits}, so the
synthesizer sees \mbox{$16{\times}16$} multipliers and infers DSPs, whereas
the other variants declare \mbox{6 bits} and the products stay in the fabric.
Therefore, HCL users still need to consider the back end, such as the DSPs inferred by the synthesiser based on data width: the language hides the syntax, not the generated hardware.

HLS, finally, solves a different problem. All HCL variants use 1096 Shift-Register LUTs (SRLs),
because their reset-less registers pack into SRLs; HLS cannot, since its
\code{DATAFLOW} processes communicate through FIFOs that carry the wide
operand structs, so it spends about \mbox{11.1k} flip-flops — three times the
pipelined variants — and the most LUTs. In exchange it gets a streaming
interface for free: the choice between an HCL and HLS is not more or less
abstraction, but structural control versus scheduling left to the tool.

Beyond QoR, each ecosystem brings its own verification solution, from fuzzing to
formal methods~\cite{spinalfuzz,chiselfv}, and with no common normalization
path, tool maturity remains a language-specific risk. Our study rests on one
kernel, one device, so the absolute numbers will move on
other targets.
\section{Conclusion}
A controlled, single-kernel comparison shows that, on a regular pipeline, the
HCLs deliver the quality of results of hand-written RTL with far less code, and
that the differences remaining between functionally identical descriptions come
from back-end arithmetic lowering and low-level typing rather than from the
algorithm. Benchmarking several languages on a single shared kernel is, in this
sense, a useful and inexpensive way to make the selection measurable. Because
QoR does not separate the languages on a regular pipeline, the decision should
follow designer background on the base language and ecosystem fit and interface needs, such as Chisel for the RISC-V
ecosystem or SpinalHDL for native streaming and bus interfaces, rather than
area or timing. We plan to extend the study in three directions: to a wider set
of kernels, including control-bound and memory-bound designs; to more devices
and an ASIC flow to validate that results are not specific application and hardware dependent.


\end{document}